\documentclass[prd,superscriptaddress,nofootinbib,notitlepage]{revtex4-1}

\usepackage{epsfig,amsmath,amssymb,slashed}
\usepackage{graphicx}
\usepackage{color}
\usepackage{siunitx}
\usepackage{color}
\usepackage{bm}
\usepackage{subfig}
\usepackage[utf8]{inputenc}
\usepackage[english]{babel}
\usepackage[T1]{fontenc}
\usepackage{amsfonts}
\usepackage{amsthm}
\usepackage{hyperref}
\usepackage{wasysym}
\usepackage{xcolor}

\newtheorem*{theorem}{Theorem}

\newcommand{\bra}[1]{\langle #1|}
\newcommand{\ket}[1]{|#1\rangle}

\newcommand{\di}{{\rm d}}
\newcommand{\ii}{i}

\def\wO{{\widehat O}}
\def\wT{{\widehat T}}

\def\wP{{\widehat P}}
\def\wJ{{\widehat J}}

\def\wpsi{{\widehat{\psi}}}
\def\wrho{{\widehat{\rho}}}
\def\wPi{{\widehat{\Pi}}}
\def\wUps{{\widehat{\Upsilon}}}

\newcommand{\tr}{{\rm tr}}  
  
\newcommand{\e}{{\rm e}}

\newcommand{\be}{\begin{equation}}
\newcommand{\ee}{\end{equation}}                                                                               
\newcommand{\bea}{\begin{eqnarray}}
\newcommand{\eea}{\end{eqnarray}}

\newcommand{\wh}{\widehat}

\begin{document}

\title{Extensivity, entropy current, area law and Unruh effect} 
\author{F. Becattini}
\affiliation{Universit\`a di Firenze and INFN Sezione di Firenze, Florence, Italy}
\author{D. Rindori}
\affiliation{Universit\`a di Firenze and INFN Sezione di Firenze, Florence, Italy}

\begin{abstract}
We present a general method to determine the entropy current of relativistic matter at local
thermodynamic equilibrium in quantum statistical mechanics. Provided that the local equilibrium 
operator is bounded from below and its lowest lying eigenvector is non-degenerate, 
it is proved that, in general, the logarithm of the partition 
function is extensive, meaning that it can be expressed as the integral over a 3D space-like hypersurface
of a vector current, and that an entropy current exists. We work out a specific calculation for 
a non-trivial case of global thermodynamic equilibrium, namely a system with constant comoving 
acceleration, whose limiting temperature is the Unruh temperature. We show that the integral
of the entropy current in the right Rindler wedge is the entanglement entropy.
\end{abstract}

\maketitle

\section{Introduction}

In recent years there has been a considerable interest on the foundations of relativistic 
hydrodynamics. One of the key quantities is the so-called \textit{entropy current} $s^\mu$, 
which is one of the postulated ingredients of Israel's formulation \cite{israel}. Therein, 
entropy current plays a very important role because, as its divergence ought to be positive, its 
form entails the constitutive equations of the conserved currents (stress-energy tensor, charged 
currents) as a function of the gradients of the intensive thermodynamic parameters. Over the last
decade, there has been a very large number of studies where the structure of the entropy current 
in relativistic hydrodynamics was involved, whose a small sample is reported in the bibliography
\cite{list1,list2,list3,list4,list5,haehl1,list6,list7}. On the other hand, there have been attempts 
\cite{kaminski} to formulate relativistic hydrodynamics without an entropy current. 

In most of these studies, the structure of the entropy current is postulated based on some classical 
form of the thermodynamics laws supplemented by more elaborate methods to include dissipative
corrections \cite{sayantani1,sayantani2}, but, strictly speaking, it is not \textit{derived}. 
The ultimate reason for the apparent insufficient definition is that the entropy current, 
unlike the stress-energy tensor or charged currents, in the familiar Quantum Field 
Theory, is not the mean value of a local operator built with quantum fields (in the 
generating functional approach, this difference can be rephrased by saying that the 
entropy current cannot be obtained by taking functional derivatives with respect to 
some external source).
\footnote{Indeed, in ref.~\cite{list6} the authors propose a method to obtain an 
entropy current from a supersymmetric generating functional, what was also the idea
of ref.~\cite{haehl2}} 
Another reason for this indeterminacy is the fact that, while in quantum statistical mechanics
the total entropy has a precise definition in terms of the density operator (von Neumann formula)
$S = -\tr (\wrho \log \wrho)$, the entropy current, which should be a more fundamental quantity
than the total entropy in a general relativistic framework, has not. 

In this work, we will show that it is indeed possible to provide a rigorous definition of the 
entropy current in quantum relativistic statistical mechanics and thereby to derive its
form in a situation of local thermodynamic equilibrium, which is the underlying assumption of
relativistic hydrodynamics. The method is based on the definition of a density operator at 
local thermodynamic equilibrium put forward in the late '70s by Zubarev and van Weert
\cite{zubarev,weert} and reworked more recently in refs.~\cite{becalocal,hongo}.
The key step to show that an entropy current exists is the proof that, in a relativistic theory, 
the logarithm of the partition function is {\em extensive}, i.e.\ it can be expressed as an
integral over a 3D space-like hypersurface of a four-vector field. We will provide, to the
best of our knowledge, the first general proof of this usually tacitly understood hypothesis
under very general conditions.

After the general method, we will present a specific, non-trivial instance of calculation of 
the entropy current which is especially interesting for it is related to the Unruh 
effect. Indeed, we will show that in the Minkowski vacuum an acceleration involves a non-vanishing 
finite entropy current in some region which, once integrated, gives rise to its total entanglement 
entropy.

\subsection*{Notation}

In this paper we adopt the natural units, with $\hbar=c=K=1$.
The Minkowskian metric tensor is $g={\rm diag}(1,-1,-1,-1)$ and for the Levi-Civita
symbol we use the convention $\varepsilon^{0123}=1$.
Operators in Hilbert space will be denoted by an upper hat, e.g.\ 
$\widehat {\sf R}$.\\
We will use the relativistic notation with repeated indices assumed to 
be saturated, and contractions of indices will be sometimes denoted with dots, e.g.\ 
$b\cdot \wh{P}=b_{\mu}\wh{P}^{\mu}$ or $\varpi :\wh{J}=\varpi_{\mu \nu}\wh{J}^{\mu \nu}$.\\
$\nabla_\mu$ denotes covariant derivative in curved space-time.

\section{Extensivity and entropy current}
\label{extensivity}

In quantum statistical mechanics, the state of a physical system is described by a density operator 
$\wrho$. At global thermodynamic equilibrium, $\wh{\rho}$ is determined, according to the maximum 
entropy principle, by maximizing the entropy $S=-\tr(\wh{\rho}\log \wh{\rho})$ with a  
set of constraints; for instance fixing the value of the energy and some possible charges.
Similarly, in a situation of \textit{local thermodynamic equilibrium} (LTE), the density operator 
$\wh{\rho}_{\rm LE}$ is determined by maximizing the entropy constrained with energy-momentum 
and charge {\em densities}. In a relativistic framework, the LTE density operator definition depends
on a 3D space-like hypersurface $\Sigma$ where the densities are given; the fully covariant expression
reads \cite{zubarev}:
\be\label{densop}
    \wh{\rho}_{\rm LE}=\frac{1}{Z_{\rm LE}}\exp \left[-\int_{\Sigma}{\rm d}\Sigma_{\mu}\left(
    \wh{T}^{\mu \nu}\beta_{\nu}-\zeta \wh{j}^{\mu}\right)\right]
\ee
where:
$$
   Z_{\rm LE}=\tr \left(\exp \left[-\int_{\Sigma}{\rm d}\Sigma_{\mu}\left(
    \wh{T}^{\mu \nu}\beta_{\nu}-\zeta \wh{j}^{\mu}\right)\right]\right)
$$
is the partition function. In the eq.~(\ref{densop}), $\wh{T}^{\mu \nu}$ is the stress-energy 
tensor, $\wh{j}^{\mu}$ is a conserved current \footnote{We do not consider
in this work anomalous currents. However, we reckon that the proposed method can 
be extended to anomalous currents as well by using the approach of ref.~\cite{becaxial}}
, $\beta_{\nu}$ is the time-like four-temperature vector, whose magnitude is the 
inverse comoving temperature:
\be\label{comtem}
  T = \frac{1}{\sqrt{\beta^2}}
\ee
while its direction defines a flow velocity \cite{becalocal}:
\be\label{velo}
  u^\mu = \frac{\beta^\mu}{\sqrt{\beta^2}},
\ee
and $\zeta$ is a scalar whose meaning is the ratio between comoving chemical 
potential and comoving temperature, i.e:
$$
  \zeta = \frac{\mu}{T}
$$

With the density operator \eqref{densop}, one can calculate the total entropy as 
a function of the four-temperature, which turns out to be:
\be\label{entropy}
     S=\log Z_{\rm LE}+\int_{\Sigma}{\rm d}\Sigma_{\mu}\left(\langle 
     \wh{T}^{\mu \nu}\rangle_{\rm LE}\beta_{\nu}-\zeta \langle \wh{j}^{\mu}\rangle_{\rm LE}\right),
\ee
where by $\langle \;\rangle_{\rm LE}$ we mean the expectation value calculated with 
$\wh{\rho}_{\rm LE}$. The expectation value of a local operator $\wO(x)$ is then 
a local intensive quantity depending only on $x$ provided that $x$ belongs to the 
hypersurface $\Sigma$ in the definition of the operator \eqref{densop}. Particularly, 
if $\Sigma$ is a hyperplane $t=const$, this means that the time component $x^0$ 
of $x$ must be equal to $t$. This condition will be henceforth understood.
For an entropy current to exist, meaning that:
\be\label{entropy2}
    S=\int_{\Sigma}{\rm d}\Sigma_{\mu}\,s^{\mu},
\ee
it can be readily seen from eq.~\eqref{entropy} that $\log Z_{\rm LE}$ must be {\em extensive}, 
namely that there must be a four-vector $\phi$, called \textit{thermodynamic potential current}, 
such that:
\begin{equation*}
    \log Z_{\rm LE}=\int_{\Sigma}{\rm d}\Sigma_{\mu}\,\phi^{\mu}
\end{equation*}
so that, comparing eqs.~\eqref{entropy} and \eqref{entropy2},
\begin{equation*}
    s^{\mu}=\phi^{\mu}+\langle \wh{T}^{\mu \nu}\rangle_{\rm LE}\beta_{\nu}- 
    \zeta \langle \wh{j}^{\mu}\rangle_{\rm LE}.
\end{equation*}
Notice that both $\phi$ and $s$ are defined up to an arbitrary four-vector tangent to the 
space-like hypersurface $\Sigma$.

The existence of the thermodynamic potential current and, as a consequence, of the
entropy current, is usually assumed. One of the goals of this work is to show that its 
existence can be proved under very general hypotheses. To begin with, let us modify 
the LTE density operator introducing a dimensionless parameter $\lambda$:
\be\label{densopl}
    \wh{\rho}_{\rm LE}(\lambda) =\frac{1}{Z_{\rm LE}(\lambda)}
    \exp \left[-\lambda \int_{\Sigma}{\rm d}\Sigma_{\mu}\left(\wh{T}^{\mu \nu}\beta_{\nu}
    -\zeta \wh{j}^{\mu}\right)\right]
\ee
so that for $\lambda=1$ we recover the actual LTE density operator. Since:
$$
    Z_{\rm LE}(\lambda) = \tr \left( 
  \exp \left[-\lambda \int_{\Sigma}{\rm d}\Sigma_{\mu}\left(\wh{T}^{\mu \nu}\, 
   \beta_{\nu}-\zeta \wh{j}^{\mu}\right)\right] \right),
$$
by taking the derivative of the trace, we come to:
$$
    \frac{\partial \log Z_{\rm LE}(\lambda)}{\partial \lambda}=
    - \int_{\Sigma}{\rm d}\Sigma_{\mu}\left(\langle \wh{T}^{\mu \nu}\rangle_{\rm LE}(\lambda)\, 
    \beta_{\nu}-\zeta \langle \wh{j}^{\mu}\rangle_{\rm LE}(\lambda)\right),
$$
and, by integrating both sides:
\be\label{zintl}
    \log Z_{\rm LE}- \log Z_{\rm LE}(\lambda_0)=-\int_{\lambda_0}^1{\rm d}\lambda 
    \int_{\Sigma}{\rm d}\Sigma_{\mu}\left(\langle \wh{T}^{\mu \nu}\rangle_{\rm LE}(\lambda)\, 
    \beta_{\nu}-\zeta \langle \wh{j}^{\mu}\rangle_{\rm LE}(\lambda)\right)
\ee
for some $\lambda_0$, being $\log Z_{\rm LE}(1)$ the actual $\log Z_{\rm LE}$. If we 
can now exchange the integrations in eq.~\eqref{zintl}, we get:
\begin{equation*}
    \log Z_{\rm LE}- \log Z_{\rm LE}(\lambda_0)=-
    \int_{\Sigma}{\rm d}\Sigma_{\mu}\, \int_{\lambda_0}^1{\rm d}\lambda \left(
    \langle \wh{T}^{\mu \nu}\rangle_{\rm LE}(\lambda)\, \beta_{\nu}
    -\zeta \langle \wh{j}^{\mu}\rangle_{\rm LE}(\lambda)\right).
\end{equation*}
Thus, if there exists a particular $\lambda_0$ such that $\log Z_{{\rm LE}}(\lambda_0)=0$, 
it is proved that $\log Z_{\rm LE}$ is extensive and, at the same time, we have a method to 
determine the thermodynamic potential current:
\begin{equation*}
    \log Z_{\rm LE} =\int_{\Sigma}{\rm d}\Sigma_{\mu}\,\phi^{\mu},\qquad
    \phi^{\mu}=-\int_{\lambda_0}^1{\rm d}\lambda \left(\langle \wh{T}^{\mu \nu}
    \rangle_{\rm LE}(\lambda)\,\beta_{\nu}-\zeta \langle \wh{j}^{\mu}\rangle_{\rm LE}(\lambda)\right).
\end{equation*}

At a glance, one would say that $\lambda=+\infty$ can make $Z_{\rm LE}(\lambda)$ very small. 
More thoroughly, one should study the spectrum of the {\em local equilibrium operator} in 
the exponent of eq.~\eqref{densop}, defined as:
\be\label{hamilt}
   \wUps \equiv \int_{\Sigma}{\rm d}\Sigma_{\mu}\left(\wh{T}^{\mu \nu}\beta_{\nu}-\zeta \wh{j}^{\mu}\right).
\ee
It is easily realized that this operator, if $\zeta=0$, is but the Hamiltonian divided by 
the temperature in the special case where $\beta= 1/T (1,{\bf 0})$.
Suppose that the local equilibrium operator $\wUps$ is bounded from below, i.e.\ there exists 
a minimum eigenvalue $\Upsilon_0$ with a corresponding eigenvector $\ket{0}$, which is 
supposedly non-degenerate. In this case, ordering the eigenvalues 
$\Upsilon_0<\Upsilon_1<\Upsilon_2\ldots$, and if the lowest eigenvector is non-degenerate
the trace can be written as:
$$
    Z_{\rm LE}(\lambda)=\tr \left(\e^{-\lambda \wUps} \right)=
    {\rm e}^{-\lambda \Upsilon_0}\left(1-{\rm e}^{-\lambda (\Upsilon_1-\Upsilon_0)}-
    {\rm e}^{-\lambda (\Upsilon_2-\Upsilon_0)}-\cdots \right),
$$
so, if $\Upsilon_0=0$ and we let $\lambda \to +\infty$, we obtain the sought solution, 
that is:
$$
    \lim_{\lambda \to +\infty}Z_{\rm LE}(\lambda)=1\qquad \implies
    \qquad \lim_{\lambda \to +\infty}\log Z_{\rm LE}(\lambda)=0.
$$
One can now take advantage of the invariance of the density operator \eqref{densop}
by subtraction of a constant number in the exponent. What we mean is that $\wrho$ 
is invariant, and entropy as well, if we replace $\wUps$ in eq.~\eqref{hamilt} with:
\begin{equation*}
    \begin{split}
        \wUps \mapsto \wUps - \Upsilon_0 &= \wUps - \bra{0} \wUps \ket{0} \\
        &= \int_{\Sigma}{\rm d}\Sigma_{\mu} \left[ \left(\wh{T}^{\mu \nu}-
        \langle 0|\wh{T}^{\mu \nu}|0\rangle\right)\beta_{\nu}-
        \zeta \left(\wh{j}^{\mu}-\langle 0|\wh{j}^{\mu}|0\rangle \right)\right]
    \end{split}
\end{equation*} 
and calculate the partition function accordingly, that is $Z'=\tr(\exp[-\wUps +\Upsilon_0])$. 
Hence, the new $\lambda$-dependent partition function reads:
$$
 Z'_{\rm LE}(\lambda)=\tr \left(
 \exp \left\{ -\lambda \int_{\Sigma}{\rm d}\Sigma_{\mu}\left[
 \left(\wh{T}^{\mu \nu}-\langle 0|\wh{T}^{\mu \nu}|0\rangle \right)\beta_{\nu}
 -\zeta \left(\wh{j}^{\mu}-\langle 0|\wh{j}^{\mu}|0\rangle \right) \right]\right\} \right)
$$
We note in passing that the subtraction of the expectation value of
$\wT^{\mu\nu}$ in the lowest lying eigenvector of $\wUps$ does not imply that 
the thereby obtained operator $\wh{T}^{\mu \nu}-\langle 0|\wh{T}^{\mu \nu}|0\rangle$ 
is the physically renormalized one. In fact, in the accelerated thermodynamic 
equilibrium case, the physical operator is obtained by subtracting the expectation
value in the Minkowski vacuum \cite{becaunruh} whereas the lowest lying eigenvector 
is the so-called Rindler vacuum, as we will see in Section~\ref{entropyacc}. 

The new partition function is such that $Z'_{\rm LE}(\infty) = 1$, and the thermodynamic 
potential current is thus given by
\be\label{thermcurr}
   \phi^{\mu}=\int_1^{+\infty} \di \lambda \left[\left(\langle\wT^{\mu\nu}\rangle_{\rm LE}(\lambda)
   - \bra{0} \wT^{\mu\nu} \ket{0}\right)\beta_{\nu}-\zeta \left(\langle \wh{j}^{\mu}\rangle_{\rm LE}(\lambda)
   -\langle 0|\wh{j}^{\mu}|0\rangle \right)\right].
\ee
Consequently, the entropy current will be:
\be\label{entcurr}
  s^{\mu} = \phi^\mu + 
  \left( \langle \wT^{\mu\nu}\rangle_{\rm LE} - \bra{0} \wT^{\mu\nu} \ket{0} \right)\beta_{\nu}
  -\zeta \left(\langle \wh{j}^{\mu}\rangle_{\rm LE}-\langle 0|\wh{j}^{\mu}|0\rangle \right).
\ee 
These formulae show that the thermodynamic potential current is obtained
by effectively integrating the temperature dependence of the currents, as $\lambda$
multiplies $\beta$ and $\zeta$. Even in the presence of first order phase transitions,
with discontinuities in the mean values of stress-energy tensor and charged current
as a function of the temperature, the integral is still feasible and makes perfect
sense.

We can finally draw an important conclusions regarding the existence of the 
entropy current. 

\begin{theorem}
If the spectrum of the local equilibrium operator \eqref{hamilt} is bounded from below, 
and if the eigenvector corresponding to its lowest eigenvalue is non-degenerate, the
logarithm of the partition function is extensive. Thus, we can obtain a thermodynamic 
potential current by integrating the difference between the expectation value of the 
stress-energy tensor and of the conserved currents and their expectation value in the
lowest lying eigenvector of the local equilibrium operator.
\end{theorem}

So, to solve the problem, one just needs to determine the mean values of the stress-energy 
tensor and of the currents. We will see how this can be accomplished in a non-trivial case
in Section~\ref{entropyacc}.

\section{Thermodynamic equilibrium with acceleration in Minkowski space-time}
\label{accetherm}

The local thermodynamic equilibrium density operator \eqref{densop} can be promoted 
to \textit{global thermodynamic equilibrium} when it becomes time-independent, or, 
in covariant language, independent of the space-like hypersurface $\Sigma$. This 
has been studied in detail elsewhere \cite{becacov}.
Global thermodynamic equilibrium occurs when $\zeta$ is constant and the four-temperature $\beta$ 
is a Killing vector, that is:
\be\label{kill}
    \nabla_{\mu}\beta_{\nu}+\nabla_{\nu}\beta_{\mu}=0.
\ee
It is readily seen from eqs.~\eqref{thermcurr} and \eqref{entcurr} that, in this case, if the 
stress-energy tensor and the currents are conserved, so are the thermodynamic potential
current and the entropy current:
$$
  \nabla_\mu \phi^\mu = 0, \qquad \nabla_\mu s^\mu = 0
$$
in agreement with the general condition that at global thermodynamic equilibrium the
entropy production rate vanishes.

In Minkowski space-time the general solution of the Killing equation~\eqref{kill} is
\begin{equation}\label{killsol}
	\beta_{\mu}=b_{\mu}+\varpi_{\mu \nu}x^{\nu},
\end{equation}
where $b$ is a constant four-vector and $\varpi$ a constant anti-symmetric tensor called 
\textit{thermal vorticity}. The latter can be expressed as the exterior derivative of the 
four-temperature, i.e.\ $\varpi_{\mu \nu}=-\frac{1}{2}(\partial_{\mu}\beta_{\nu}-\partial_{\nu}\beta_{\mu})$. 

By using the eq.~\eqref{killsol}, one can obtain the general form of the density
operator \eqref{densop} in Minkowski space-time, that is:
\be\label{densop2}
\wh{\rho}=\frac{1}{Z}\exp \left[-b_{\mu}\wP^{\mu} + \frac{1}{2} \varpi_{\mu \nu}\wJ^{\mu \nu}+\zeta \wh{Q} \right]
\ee
where $\wP$ is the four-momentum operator, $\wJ$ the boost-angular momentum operator and $\wh{Q}$ the 
conserved charge associated to the current $\wh{j}$. Among the various solutions, a noteworthy one 
is the {\em pure acceleration} one
\be\label{pureacc}
	b_{\mu}=\frac{1}{T_0}(1,{\bf 0}),\qquad \varpi_{\mu \nu}=
	\frac{a}{T_0}(g_{0\nu}g_{3\mu}-g_{3\nu}g_{0\mu})
\ee
with constant parameters $a$ and $T_0$. This case has been studied in detail in ref.~\cite{becaunruh}, 
and corresponds to a fluid with four-temperature field:
\be\label{flow}
	\beta^{\mu}=\frac{a}{T_0}\left(z',0,0,t\right)
\ee
where $z'\equiv z+1/a$, and a four-acceleration field:
\be\label{accel}
   A^\mu = \frac{1}{z'^2-t^2}  \left(t,0,0,z'\right).
\ee   
It is readily found that the $\beta$ field \eqref{flow} is time-like for $|z'|>t$, light-like 
for $|z'|=t$ and space-like for $|z'|<t$. Hence, the hypersurfaces 
$|z'|=t$ are two Killing horizons for $\beta$ and they break the space-time into four different 
regions: the region $|t|<z'$ is called \textit{right Rindler wedge} (RRW), the region $|t|<-z'$ 
is called \textit{left Rindler wedge} (LRW), while the regions $|t|>|z'|$ are not of interest
here. The proper temperature \eqref{comtem} in the RRW is singular on the light-cone boundary:
$$
  T = \frac{T_0}{a} \frac{1}{\sqrt{z'^2-t^2}}.
$$

It is also very useful to decompose the thermal vorticity tensor $\varpi$ into two space-like
vector fields $\alpha$ and $w$ \cite{buzzegoli} by projecting onto the velocity field \eqref{velo}.
In the global equilibrium case, they turn out to be parallel to 
the four-acceleration and the kinematic vorticity respectively, and both orthogonal to $u$ or $\beta$. In 
the pure acceleration case, it turns out that the kinematic vorticity vanishes and we are left 
with:
\be\label{thvort}
    \varpi_{\mu \nu} = \alpha_{\mu}u_{\nu}-\alpha_{\nu}u_{\mu},
\ee
where $\alpha^\mu = A^\mu/T$ is the four-acceleration divided by the comoving temperature.
By using \eqref{accel}, \eqref{flow} and \eqref{comtem} to calculate $T$, it turns out that:
\be\label{alpha2}
  \alpha^2 = \frac{A^2}{T^2} = \beta^2 A^2 = -\frac{a^2}{T_0^2},
\ee
that is, $\alpha^2$ is constant.

In the pure acceleration case \eqref{pureacc} with $\zeta=0$, the density operator \eqref{densop2} 
becomes:
\be\label{densopacc}
\wh{\rho}=\frac{1}{Z}\exp \left[-\frac{\wh{H}}{T_0}+\frac{a}{T_0}\wh{K}_z\right]
\ee
with $\wh{K}_z=\wh{J}_{30}$ the boost generator along the $z$ direction. The operator 
$\wh{H}-a\wh{K}_z$ is stationary and can be regarded as the generator of translations 
along the flow lines of \eqref{flow} \cite{leinaas}.

The peculiarity of the four-temperature field \eqref{flow} is that it altogether vanishes 
on the 2D surface $t=0, z'=0$, what makes it possible to factorize the density operator 
into two commuting operators involving the quantum field degrees of freedom on either side 
of $z'=0$, i.e.
\begin{equation}\label{densopfact}
    \begin{split}
        \wh{\rho}=\wh{\rho}_R\otimes \wh{\rho}_L,&\qquad [\wh{\rho}_R,\wh{\rho}_L]=0\\
        \wh{\rho}_R = \frac{1}{Z_R} \exp \left[-\frac{\wPi_R}{T_0}\right],
        &\qquad \wh{\rho}_L = \frac{1}{Z_L} \exp \left[\frac{\wPi_L}{T_0}\right]
    \end{split}
\end{equation}
with:
\be\label{pi}
   \wPi_{R,L} \equiv T_0 \int_{z'\lessgtr 0}{\rm d}\Sigma_{\mu}\,
    \wh{T}^{\mu \nu}\beta_{\nu}
\ee
being the generator of translations along the flow lines, playing the role of the
Hamiltonian. The factorization implies that, if $\wO(x)$ is a local operator with $x$ in 
the RRW, its expectation value is independent of the field operators in the LRW.

Quantum field equations of motion, such as Klein-Gordon, are solved in the RRW and LRW separately
by introducing proper hyperbolic coordinates, the \textit{Rindler coordinates} $(\tau,x,y,\xi)$,
where the ``transverse'' coordinates $\mathbf{x}_T\equiv (x,y)$ are the same as Minkowski's, and 
$(\tau,\xi)$ are related to $(t,z')$ by
\begin{equation*}
    \tau \equiv \frac{1}{2a}\log \left(\frac{z'+t}{z'-t}\right),\qquad \xi \equiv 
    \frac{1}{2a}\log \left[a^2\left({z'}^2-t^2\right)\right]
\end{equation*}
in the RRW. Inverting and plugging them into the Klein-Gordon equation, one finds the
positive-frequency modes \cite{crispino}
\begin{equation*}
    u_{\omega,\mathbf{k}_T}(\tau,\xi,\mathbf{x}_T)\equiv \sqrt{\frac{1}{4\pi^4a}
    \sinh \left(\frac{\pi \omega}{a}\right)}{\rm K}_{i\frac{\omega}{a}}
    \left(\frac{m_T{\rm e}^{a\xi}}{a}\right){\rm e}^{-i(\omega \tau-\mathbf{k}_T\cdot \mathbf{x}_T)},
\end{equation*}
where $\omega$ is a positive real number, $\mathbf{k}_T\equiv (k_x,k_y)$ is the ``transverse'' momentum, 
${\rm K}$'s are the modified Bessel functions and $m_T^2\equiv \omega^2-\mathbf{k}_T^2$.
The real scalar field can thus be expanded as:
\be\label{field}
    \wh{\psi}(\tau,\xi,\mathbf{x}_T)=\int_0^{+\infty}{\rm d}\omega \int_{\mathbb{R}^2}{\rm d}^2k_T\left(u_{\omega,\mathbf{k}_T}\wh{a}^R_{\omega,\mathbf{k}_T}+
    u^*_{\omega,\mathbf{k}_T}\wh{a}^{R\dagger}_{\omega,\mathbf{k}_T}\right),
\ee
where $\wh{a}^{R\dagger}$ and $\wh{a}^R$ are the creation 
and annihilation operators respectively, and they satisfy the usual commutation relations.
A similar field expansion holds in the LRW, with the important difference that the role of the 
creation and annihilation operators is interchanged, as a consequence of the fact that the boost 
generator, which plays the role of time translations generator, is past-oriented therein.
The ``Hamiltonians'' $\wPi$'s turn out to be \cite{becaunruh}:
\be\label{pirl}
    \wPi_{R,L}= \int_0^{+\infty}{\rm d}\omega \int_{\mathbb{R}^2}{\rm d}^2k_T\,
    \wh{a}^{\dagger R,L}_{\omega,\mathbf{k}_T}\, \wh{a}^{R,L}_{\omega,\mathbf{k}_T}.
\ee
The vacuum in the RRW is obtained requiring it to be annihilated by all the operators 
$\wh{a}^R$, and it is denoted as $\ket{0_R}$. Similarly, 
the vacuum $\ket{0_L}$ in the LRW is the state annihilated by all $\wh{a}^L$, and the overall
vacuum state $\ket{0}_R\equiv \ket{0_R}\otimes \ket{0_L}$ is the so-called \textit{Rindler vacuum}.
As first pointed out by Fulling in \cite{fulling},  the Rindler vacuum does not coincide with
the Minkowski vacuum. In fact, the Minkowski vacuum is a thermal state of free bosons with
temperature $a/2\pi$, the well-known Unruh effect \cite{unruh}.

The thermal expectation values of relevant physical quantities in the RRW, for a free field, 
with the density operator \eqref{densopfact} can be obtained once the thermal expectation values 
of products of creation and annihilation operators are known. In \cite{becaunruh} the following 
expressions were found:
\begin{subequations}\label{aacroce}
    \begin{align}
        \left\langle \wh{a}^{R\dagger}_{\omega,\mathbf{k}_T}\wh{a}^R_{\omega',\mathbf{k}'_T}\right\rangle 
        =&\frac{1}{{\rm e}^{\omega/T_0}-1}\delta(\omega-\omega')\,\delta^2(\mathbf{k}_T-\mathbf{k}'_T)\label{aacroce1}\\
        \left\langle \wh{a}^R_{\omega,\mathbf{k}_T}\wh{a}^{R\dagger}_{\omega',\mathbf{k}'_T}\right\rangle =&\left(\frac{1}{{\rm e}^{\omega/T_0}-1}+1\right)\delta(\omega-\omega')\,\delta^2(\mathbf{k}_T-\mathbf{k}'_T)
        \label{aacroce2}\\
        \left\langle \wh{a}^R_{\omega,\mathbf{k}_T}\wh{a}^R_{\omega',\mathbf{k}'_T}\right\rangle=&\left\langle \wh{a}^{R\dagger}_{\omega,\mathbf{k}_T}\wh{a}^{R\dagger}_{\omega',\mathbf{k}'_T}\right\rangle=0
        \label{aacroce3},
    \end{align}
\end{subequations}
and it is shown that normal ordering with respect to the Rindler vacuum corresponds to neglect 
the $+1$ in \eqref{aacroce2}, which arises from the commutation relations of creation and annihilation 
operators. From formulae \eqref{aacroce}, the following normally ordered expression can be calculated:
\begin{equation*}
    \left\langle {:\wh{\psi}^2:}_R\right\rangle=\int_0^{+\infty}{\rm d}\omega \int_{\mathbb{R}^2}{\rm d}^2k_T\,\left|u_{\omega,\mathbf{k}_T}\right|^2\frac{2}{{\rm e}^{\omega/T_0}-1},
\end{equation*}
where the subscript $R$ stands for the normal ordering of Rindler creation and annihilation 
operators and corresponds to the subtraction of the expectation values in the Rindler vacuum.
For a massless field, the above integration can be carried out analytically, yielding:
\begin{equation*}
    \left\langle {:\wh{\psi}^2:}_R\right\rangle=\frac{T_0^2}{12} \frac{1}{a^2(z'^2-t^2)} = 
    \frac{1}{12\beta^2},
\end{equation*}
where the eq.~\eqref{flow} has been used. Another useful expression which was obtained in 
ref.~\cite{becaunruh} is:
\begin{equation}\label{eq14}
    \left\langle {:u\cdot \partial \wh{\psi}\,u\cdot \partial \wh{\psi}:}_R\right\rangle
     =\frac{\pi^2}{30\beta^4}.
\end{equation}
%

\section{Entropy current for a free scalar field}
\label{entropyacc}

We are now in a position to calculate the thermodynamic potential current and the entropy current
for a relativistic fluid at global thermodynamic equilibrium with acceleration in the RRW.

The basic ingredient to determine the entropy current, according to eqs.~\eqref{thermcurr} and \eqref{entcurr},
is the mean value of the stress-energy tensor. The general form of the mean value of a 
symmetric rank-2 tensor at global thermodynamic equilibrium with acceleration is constrained by 
the symmetries of the density operator \eqref{densopacc}, as well as by the parameters at 
our disposal. In general, looking at the equation \eqref{densop2}, the mean value of any 
local operator $\wO$ can be a function of $b,\varpi,x$; however, the dependence on $x$ is 
constrained by the form of the density operator.
Denoting by $\widehat{\sf T}(x)$ the translation operator $\exp[\ii x \cdot \wP]$, we have: 
\begin{equation}\label{xdep}
    \begin{split}
        \langle \wO(x)\rangle &= \langle \widehat{\sf T}(x)\, \wO(0)\, \widehat{\sf T}^{-1}(x) \rangle
        = \tr \left( \widehat{\sf T}^{-1}(x)\, \wrho \; \widehat{\sf T}(x)\, \wO(0) \right) \\
        &= \frac{1}{Z} \tr \left( \e^{-b \cdot \wP + \frac{1}{2} \varpi: \widehat{\sf T}^{-1}(x) \wJ 
        \widehat{\sf T}(x)} \wO(0) \right) = \frac{1}{Z} \tr \left( \e^{-\beta(x) \cdot \wP + 
    \frac{1}{2} \varpi:\wJ} \wO(0) \right) = \langle \wO(0)\rangle_{\beta(x)}
    \end{split}
\end{equation}
where we have used the known relation from Poincar\'e algebra:
$$
  \widehat{\sf T}^{-1}(x) \wJ_{\mu\nu} \widehat{\sf T}(x) =  \wJ_{\mu\nu} -
  x_\nu \wP_\mu + x_\mu \wP_\nu
$$
and the eq.~\eqref{killsol}. The eq.~\eqref{xdep} implies that the expectation value 
of {\em any} local operator at global thermodynamic equilibrium depends on the space-time point 
only through the four-temperature vector field $\beta(x)$. Therefore, we can build up the 
most general expectation value of tensor fields of any rank just by combining $\beta$, the 
constant anti-symmetric tensor $\varpi$ and the metric tensor $g$. Note that $\partial_\mu \beta_\nu = \varpi_{\nu\mu}$ and $\partial^2 \beta = 0$, so that derivatives of $\beta$ cannot enter as 
independent variables.\\
Likewise, any scalar $F$ mean value of a local operator or derived from it, at global 
thermodynamic equilibrium with acceleration can only be a function 
of the two scalars formed with $\beta$ and $\varpi$, that is $\beta^2 = 1/T^2$ and $\alpha^2$, 
taking into account that $\alpha\cdot\beta = 0$. Therefore a rank-2 symmetric tensor such as 
the stress-energy tensor, must have the following form:
\begin{equation}\label{set}
    \langle \wh{T}^{\mu \nu}\rangle=F_1\beta^{\mu}\beta^{\nu}+F_2g^{\mu \nu}+F_3\alpha^{\mu}\alpha^{\nu} 
    + F_4\left(\alpha^{\mu}\beta^{\nu}+\alpha^{\nu}\beta^{\mu}\right)
\end{equation}
with $F_i=F_i(\beta^2,\alpha^2)$. Furthermore, if the Hamiltonian $\widehat{H}$ in eq.~\eqref{densopacc}
is invariant under time-reversal, so is the density operator itself; as a consequence, the mixed 
components of the stress-energy tensor at $t=0$ must vanish, that is  $\langle \wT^{0i}(0,{\bf x})\rangle=0$. 
Since at $t=0$ we have $\alpha^i \ne 0$ and $\beta^0 \ne 0$, the term proportional to $F_4$ in
eq.~\eqref{set} breaks time-reversal invariance, and $F_4$ must then vanish. Similarly, it can
be shown that the most general expression of the expectation value of a vector field should be of
the simple form:
$$
   \langle \wh{V}^\mu \rangle = F(\beta^2,\alpha^2) \beta^\mu
$$
being the terms linear in $\alpha^\mu$ forbidden by time-reversal invariance.

According to the formulae \eqref{thermcurr} and \eqref{entcurr}, one also needs to determine the
expectation value in the eigenvector of $\wUps$ with the lowest eigenvalue. It is easy to realize
that, looking at the equations \eqref{densopfact} and \eqref{pirl}, for a free field this eigenvector 
is just the Rindler vacuum $\ket{0}_R = \ket{0_R}\otimes \ket{0_L}$, whose eigenvalue is zero. 
Then, since it is not degenerate, the Rindler vacuum expectation value of any operator must have the same 
symmetries as its expectation value with the density operator \eqref{densopfact}, and we can thus 
write:
\be\label{set2}
   \langle : \wT^{\mu\nu} :_R \rangle \equiv
   \langle \wh{T}^{\mu \nu}\rangle - {}_R\bra{0} \wT^{\mu\nu} \ket{0}_R =  
   F_1\beta^{\mu}\beta^{\nu}+F_2g^{\mu \nu}+F_3\alpha^{\mu}\alpha^{\nu}.
\ee
Contracting \eqref{set2} with the four-temperature twice, we obtain
\begin{align*}
        \langle : \wT^{\mu\nu} :_R \rangle \beta_{\nu} = &
        \left(F_1\beta^2+F_2\right)\beta^{\mu}\\ 
        F_1 \beta^2+F_2 = & \langle : \wT^{\mu\nu} :_R \rangle
        \frac{\beta_{\mu}\beta_{\nu}}{\beta^2}=
        \langle : \wT^{\mu\nu} :_R \rangle u_{\mu}u_{\nu}=\rho
\end{align*}
$\rho$ being the energy density. In summary, we have
$$
    \langle : \wT^{\mu\nu} :_R \rangle \beta_{\nu} =\rho \beta^{\mu}
$$
and, as a consequence:
\be\label{thermcurr2}
    \phi^{\mu} = \beta^\mu \int_1^{+\infty}{\rm d}\lambda \,\rho(\lambda)
\ee
with $\rho(\lambda)$ obtained from $\langle {:\wh{T}^{\mu \nu}:}_R\rangle(\lambda)$.

For the free real scalar field \eqref{field} we can obtain the canonical stress-energy tensor
from the Lagrangian density:
$$
    {\cal L}=\frac{1}{2}\partial_{\mu}\wh{\psi}\, \partial^{\mu}\wh{\psi}-
    \frac{1}{2}m^2\wh{\psi}^2,
$$
that is:
\be\label{canonical}
  \wh{T}^{\mu \nu}_{\rm CAN}
  =\partial^{\mu}\wh{\psi}\, \partial^{\nu}\wh{\psi}-\frac{1}{4}\Box \wh{\psi}^2g^{\mu \nu},
\ee  
where we have used the equations of motion $(\Box + m^2) \wpsi = 0$. The energy density 
thus reads:
\begin{equation}\label{eq11}
    \rho=\langle : \wh{T}^{\mu \nu} :_R \rangle u_{\mu}u_{\nu}
    =\left\langle : u\cdot \partial \wh{\psi}\, u\cdot \partial \wh{\psi} :_R \right\rangle-
    \frac{1}{4}\Box \left\langle :\wh{\psi}^2 :_R \right\rangle.
\end{equation}

By using the $\beta$ expression in \eqref{flow} we obtain
$$
  \Box \left\langle {:\wh{\psi}^2:}_R\right\rangle=\frac{\alpha^2}{3\beta^4},
$$
which, plugged into \eqref{eq11} together with \eqref{eq14}, gives the energy 
density of a free real scalar field in the RRW
\begin{equation}\label{endens}
    \rho=\frac{\pi^2}{30\beta^4}-\frac{\alpha^2}{12\beta^4}.
\end{equation}
This expression is precisely what was found in ref.~\cite{becagrossi,buzzegoli} with a 
perturbative expansion of the density operator \eqref{densopacc} in $\alpha$ at order 
$\alpha^2$. Hence, we found out that the perturbative series for the real scalar field
is simply a polynomial in $\alpha$ of order 2.

We now need the function $\rho(\lambda)$ to calculate the thermodynamic potential current.
From eqs.~\eqref{densopl} and \eqref{densopacc} it turns out that the introduction of the 
dimensionless parameter $\lambda$ corresponds to the rescaling $T_0\mapsto T_0/\lambda$. 
To make the full dependence of $\rho$ on $T_0$ apparent, it is convenient to introduce the 
Killing vector $\gamma \equiv T_0\beta$ which is independent of $T_0$ (see eq.~\eqref{flow}) 
and, taking into account \eqref{alpha2}, write the above energy density as:
$$
 \rho =\frac{\pi^2}{30\gamma^4}T_0^4 + \frac{a^2}{12\gamma^4} T_0^2.
$$
Rescaling $T_0$, we readily obtain:
$$
  \rho(\lambda)=\frac{\pi^2}{30\gamma^4}\frac{T_0^4}{\lambda^4}+
  \frac{a^2}{12\gamma^4}\frac{T_0^2}{\lambda^2} = \frac{\pi^2}{30\beta^4}\frac{1}{\lambda^4}-
  \frac{\alpha^2}{12\beta^4}\frac{1}{\lambda^2}.
$$
Plugging it into \eqref{thermcurr2} we get the thermodynamic potential current:
$$
  \phi^{\mu}=\left(\frac{\pi^2}{90\beta^4}-\frac{\alpha^2}{12\beta^4}\right)\beta^{\mu}
$$
and, consequently, the entropy current in the RRW
\be\label{entcurr2}
  s^{\mu}=\left(\frac{2\pi^2}{45\beta^4}-\frac{\alpha^2}{6\beta^4}\right)\beta^{\mu}.
\ee

It should be pointed out that the above formulae depend on the stress-energy quantum
operator. Indeed, for the improved stress-energy tensor, which is traceless for a 
massless field:
\be\label{improved}
  \wT^{\mu\nu}_{\rm IMP} = \wT^{\mu\nu}_{\rm CAN} - \frac{1}{6} (\partial^\mu \partial^\nu
  - g^{\mu\nu} \Box) \wpsi^2,
\ee
we obtain a different expression of the energy density at equilibrium. This is an expected
feature of thermodynamic equilibrium with rotation or acceleration, as was extensively 
discussed in ref.~\cite{becatinti}. Indeed, the additional term to the energy density 
pertaining to the canonical stress-energy tensor in eq.~\eqref{improved} turns out to be:
$$
- \frac{1}{6} (u^\mu u^\nu \partial_\mu \partial_\nu - \Box) \left\langle {:\wh{\psi}^2:}_R \right\rangle
  =- \frac{1}{6} (u^\mu u^\nu \partial_\mu \partial_\nu - \Box) \frac{1}{12 \beta^2} = \frac{\alpha^2}{12\beta^4}
$$
as it can be shown by using eq.~\eqref{flow}. Thus, adding the above contribution to 
eq.~\eqref{endens}, we find:
$$
 \rho_{\rm IMP} = \frac{\pi^2}{30\beta^4}
$$
that is, the energy density calculated with the improved stress-energy tensor for the 
massless free real scalar field depends only on $\beta^2$ and not on $\alpha^2$, which is a 
somewhat surprising feature. Likewise, the entropy current gets modified and one is left
with only the first term of eq.~\eqref{entcurr2}:
\be\label{entcurr3}
  s^{\mu}_{\rm IMP} = \frac{2\pi^2}{45\beta^4} \beta^{\mu}.
\ee
%
\begin{figure}
	\includegraphics[scale=0.9]{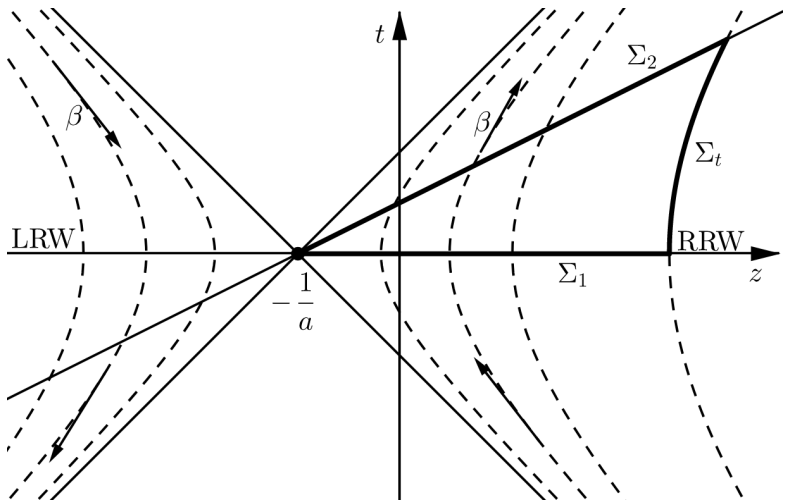}
	\caption{Two-dimensional section of the Minkowski space-time with the Killing field
	$\beta$ in eq.~\eqref{flow} splitting the plane $tz$ into the Rindler wedges bounded
	by the light-cone in $z=-1/a$. The integrals of conserved currents on the space-like 
	hypersurfaces $\Sigma_1$ and $\Sigma_2$ in the RRW are the same because of the Gauss
	theorem and taking into account that the time-like hyperbolic boundary $\Sigma_t$ yields
	no contribution, being $\beta$ perpendicular to its normal unit vector.}
	\label{figure}
\end{figure}

\section{Entanglement entropy, area law and Unruh effect}
\label{entanglement}

The eq.~\eqref{entcurr2} is the entropy current in the RRW, therefore, its integral on a space-like 
hypersurface having as boundary the 2D surface $z'=0,t=0$ (see fig.~\ref{figure}) is 
\be\label{enteng}
  S_R = \int_{z'>0} \di \Sigma_\mu \, s^\mu = - \tr_R (\wrho_R \log \wrho_R) 
\ee
according to eqs.~\eqref{densopfact} and \eqref{pi} and by means of the previous construction of the
current. As the density operator is factorized, this entropy is also the entanglement entropy
obtained by tracing out the field degrees of freedom in the LRW. At global thermodynamic equilibrium
we have $\partial_\mu s^\mu = 0$, and the entropy \eqref{enteng} can be calculated on any 
space-like hypersurface provided that the boundary flux vanishes. Indeed, this is the case
for the RRW, as the time-like boundary is tangent to the entropy current (see fig.~\ref{figure}).

A straightforward calculation of the entanglement entropy on the hypersurface $t=0$ with
the canonical entropy current \eqref{entcurr2} yields:
\be\label{enteng2}
    S_R = \int_{\mathbb{R}^2} \di x \, \di y \left(\frac{2\pi^2}{45}-\frac{\alpha^2}{6}\right)
    \frac{T_0^3}{a^3} \lim_{z'\to 0}\frac{1}{2{z'}^2} .
\ee    
Thus, the entropy turns out to be proportional to the area of the 2D boundary surface 
separating the RRW from the LRW but with a divergent constant, which is owing to the 
fact that the comoving temperature $T$ diverges for $z'=0$. This result is in full agreement
with that of Bombelli \textit{et al.} and Srednicki \cite{bombelli,srednicki}.

The same result can be obtained with a more general and more elegant derivation which applies
to general space-times. As the entropy current $s$ is divergenceless, $\nabla_\mu s^\mu =0$, it can 
be expressed it as the Hodge dual of an exact 3-form. If the domain is topologically contractible 
(so is the RRW), this form in turn can be expressed as the 
exterior derivative of a 2-form \cite{wald,oz,padmanabhan1,padmanabhan2}, which eventually amounts to state that 
the original vector field can be written as the divergence of an anti-symmetric tensor field:
$$
     s^\mu = \nabla_\nu \varsigma^{\mu\nu}
$$
whence, because of the Stokes' theorem:
\be\label{area}
  S=\int_{\Sigma} \di\Sigma_{\mu} \; s^\mu = \int_{\Sigma} \di\Sigma_{\mu} \; 
  \nabla_\nu \varsigma^{\mu\nu}
   = \frac{1}{2}\int_{\partial \Sigma} \di\tilde{S}_{\mu \nu} \,\varsigma^{\mu \nu}
   = - \frac{1}{4} \int_{\partial \Sigma} \di S^{\rho\sigma} \; \sqrt{-g} \, 
    \epsilon_{\mu\nu\rho\sigma} \,\varsigma^{\mu \nu}
\ee
where $\di S^{\rho\sigma}$ is the measure of the 2D boundary surface $\partial\Sigma$. Hence,
the total entropy is expressed as a surface integral of a potential of a conserved current
\cite{wald}. For our specific problem of equilibrium with acceleration, the general expression
of the potential turns out to be:
\be\label{potential}
    \varsigma^{\mu\nu}=\frac{s}{2\alpha^2}\left(\beta^{\mu}\alpha^{\nu}
     -\beta^{\nu}\alpha^{\mu}\right),
\ee
where $s\equiv s^\mu u_\mu$ is the \textit{entropy density}. Plugging eq.~\eqref{potential}
into eq.~\eqref{area}:
$$
   S = -\frac{1}{4} \int_{\partial\Sigma} \di S^{\rho\sigma} \; \sqrt{-g} 
    \frac{s}{2\alpha^2}\, \epsilon_{\mu\nu\rho\sigma} 
    \left(\beta^{\mu}\alpha^{\nu} -\beta^{\nu}\alpha^{\mu}\right).
$$
The boundary of the hypersurface $t=0$ is the plane $t=0,z'=0$ and the plane $t=0,z'=+\infty$.
In the latter, the integrand vanishes for $s \propto z^{\prime-3}$, $\alpha^2$ is constant
and $\beta^0 \alpha^3 \propto z'$. We are thus left with 
the $xy$ plane and, taking into account that $\rho,\sigma$ indices can only take on values 
$1,2$ and of the dependence of $\beta$ and $\alpha$ on $(z',t)$, we end up with the same 
eq.~\eqref{enteng2}.

A remarkable consequence of the entropy current method is the determination of the entanglement
entropy in the Minkowski vacuum, when the state of the system is pure $\wrho = \ket{0_M}\bra{0_M}$.
It is well known that \cite{crispino}:
$$ 
  \wrho_R = \tr_L (\ket{0_M}\bra{0_M}) = \frac{1}{Z_R} \exp \left[ -\frac{2\pi}{a} \wPi_R \right]
$$
that is, the Minkowski vacuum for a system with acceleration $a$ corresponds, in the RRW, to a mixed 
state with density operator \eqref{densopfact} with $T_0 = a/2\pi$, 
which is in essence the content of the Unruh effect. It was observed in
\cite{becaunruh} that, from a statistical thermodynamics viewpoint, this corresponds to a limiting 
comoving temperature of $T_U = |A|/2\pi$, where $|A|$ is the magnitude of the four-acceleration field.
Because of \eqref{alpha2}, we thus have an upper bound for $|\alpha^2| = (2 \pi)^2$ in the Minkowski
vacuum, and so eq.~\eqref{entcurr2} becomes:
$$
   s^\mu = \frac{32\pi^2}{45} T^3_U u^\mu,
$$
while for the eq.~\eqref{entcurr3}
$$
   s^\mu_{\rm IMP} = \frac{2\pi^2}{45} T^3_U u^\mu.
$$
which mean that we have a non-vanishing entropy current in the Minkowski vacuum, which is owing
to having traced out the field degrees of freedom in the LRW.
Remarkably, the above expressions differ by a factor 16, which is apparently an unexpected and 
odd feature. 
Yet, as it has been mentioned, at global thermodynamic equilibrium the mean value of the
stress-energy tensor does depend on the specific quantum operator (in the case at hand, either
\eqref{canonical} or \eqref{improved}) and the entropy current as well. On the other hand, the 
total integrals like $\wP^{\mu}$ and $\wJ^{\mu\nu}$ should not depend on it (see the discussion 
in ref.~\cite{becatinti}), and, as a consequence, the entanglement entropy should also be independent 
because $\wrho_R$ can be written as a trace over the field degrees of freedom of a density operator 
which is function of the Poincar\'e generators (see eq.~\eqref{densopacc}):
$$
   \wrho_R = \tr_L (\wrho) = \frac{1}{Z} \tr_L \left( 
    \exp \left[-\frac{\wh{H}}{T_0}+\frac{a}{T_0}\wh{K}_z\right] \right).
$$
Nevertheless, the expressions of $\wPi_R$ and $\wPi_L$ (see eqs.~\eqref{densopfact}, \eqref{pi}) 
may inherit a dependence on the quantum stress-energy tensor because of the truncation 
at $z'=0$. This issue will be the subject of further investigation.

\section{Summary and outlook}

In summary, we have presented the condition of existence of an entropy current and a general 
method to calculate it. An entropy current can be obtained if the spectrum of the local equilibrium 
operator, which boils down to the Hamiltonian multiplied by $1/T$ in the simplest case of global 
homogeneous equilibrium, is bounded from below.
We have applied our method to the case of a fluid with a comoving acceleration of constant 
magnitude, a known instance of non-trivial global thermodynamic equilibrium in Minkowski space-time. 
We have also shown its connection to the entanglement entropy in the vacuum, and its relation with 
the Unruh effect. Furthermore, we have shown that, at least at global equilibrium, the total entropy 
can be expressed as a surface integral, in agreement with \cite{wald}. We expect this method to be 
applicable to other problems where the total entropy has to be determined, like e.g.\ relativistic 
hydrodynamics or thermodynamic equilibrium in general curved space-time. 

The entropy current is expectedly dependent on the specific form of the stress-energy tensor 
operator. Besides, in the case of the free scalar field, the canonical (minimal
coupling) and the improved stress-energy tensor (conformal coupling) seem to imply two different
results for the total entanglement entropy, even though infinite. This is a subject for future
studies.

\section*{Acknowledgments}

We acknowledge useful discussions with D. Seminara.


\appendix

\section{Calculation of the entropy potential}

The search of a potential for the entropy current uses the same method as for the stress-energy
tensor. To form an anti-symmetric tensor we can just use the four-vectors $\beta$ and $\alpha$,
hence the only possible combination is:
$$
  \alpha^{\mu}\beta^{\nu}-\alpha^{\nu}\beta^{\mu}.
$$
This is, in turn, just proportional to the thermal vorticity, according to eq.~\eqref{thvort},
so, we can write the general form of the potential like this:
\be\label{genpot}
  \varsigma^{\mu\nu} = G(\beta^2,\alpha^2) \varpi^{\mu\nu}
\ee
being $G$ a general scalar function such that:
$$
  s^\mu = \partial_\nu \left( G \varpi^{\mu\nu} \right).
$$
By introducing the proper entropy density $s$ such that $ s^\mu u_\mu = s$, we have:
$$
 s = s^\mu u_\mu = \frac{1}{\sqrt{\beta^2}} s^\mu \beta_\mu = \frac{1}{\sqrt{\beta^2}} 
 \beta_\mu \varpi^{\mu\nu} \partial_\nu G
$$
as $\varpi$ is constant. Now, $\varpi^{\mu\nu} \beta_\mu = - \sqrt{\beta^2} \alpha^\nu$
from \eqref{thvort} and so:
\be\label{entdens}
  s = - \alpha^\nu \partial_\nu G = -\alpha^\nu \partial_\nu \beta^2 
   \frac{\partial G}{\partial \beta^2}
\ee
because $\alpha^2$ is a constant in the pure acceleration case. The Killing equation 
\eqref{kill} implies \cite{buzzegoli}:
$$
  \partial_\nu \beta^2 = - 2 \sqrt{\beta^2} \alpha_\nu
$$
so that the \eqref{entdens} becomes:
\begin{equation*}
  s = 2 \alpha^2  \sqrt{\beta^2} \frac{\partial G}{\partial \beta^2}
\end{equation*}
whose solution is:
$$
  G = \int \di \beta^2 \; \frac{s}{2\alpha^2} \frac{1}{\sqrt{\beta^2}}.
$$

For the massless case, we have $s = C(\alpha)^2 / \sqrt{\beta^2}^3$ from \eqref{entcurr2}, hence:
$$
  G = \frac{C(\alpha^2)}{2 \alpha^2} \int \di \beta^2 \frac{1}{\beta^4} 
    = - \frac{C(\alpha^2)}{2 \alpha^2} \frac{1}{\beta^2} = - s \frac{2 \alpha^2}{\sqrt{\beta^2}}.
$$    
Plugging the above result into the \eqref{genpot} and using the \eqref{thvort} we obtain:
$$
  \varsigma^{\mu\nu} = - \frac{s}{2 \alpha^2} \sqrt{\beta^2} \varpi^{\mu\nu} = -
   \frac{s}{2 \alpha^2} \sqrt{\beta^2} (\alpha^\mu u^\nu - \alpha^\nu u^\mu) = 
   \frac{s}{2 \alpha^2} (\alpha^\nu \beta^\mu - \alpha^\mu \beta^\nu)
$$
which is precisely \eqref{potential}.

\end{document}